\documentclass{article}

\usepackage[T1]{fontenc}
\usepackage{url}
\usepackage[utf8]{inputenc} 
\usepackage{amsmath}
\usepackage{graphicx}
\usepackage[usenames,dvipsnames]{color}
\begin{document}
\title{A simple model of knowledge percolation}
\author{Franco Bagnoli$^{1,2}$ and Guido de Bonfioli Cavalcabo'$^{1}$ \\
    (1) Department of Physics and Astronomy and CSDC, \\
    University of Florence,\\
    via G. Sansone 1 - 50019 Sesto Fiorentino (Italy)\\
    (2) INFN, sez.\ Firenze.   \\ 
       \texttt{franco.bagnoli@unifi.it}, \\
    \texttt{guido.debonfiolicavalcabo@stud.unifi.it} 
}

\maketitle             

\begin{abstract}
We investigate how knowledge percolates and clusters in a given knowledge space.
We introduce a simple model of knowledge organization in which each contribution spans a certain number of items. If this contribution overlaps with others above a certain threshold, they form a cluster. A contribution can also merge clusters together. We study the growth of global knowledge and the cluster dynamics, both showing a nontrivial  behavior.
\end{abstract}

\section{Introduction}

Knowledge is the set of ideas, emotions, beliefs and experiences, such as facts (descriptive knowledge), skills (procedural knowledge), or objects (acquaintance knowledge) owned by an individual or shared across collaborating individuals~\cite{Kumbhar2012,knowledge_evolution}. It can be roughly seen as a set of concepts linked by some relationship (e.g. derivations linked to prerequisites or axioms to form theorems). 
The set of knowledge items that are connected by a path of links  can be denoted as a cluster of knowledge. 

A good representation of this description is a network~\cite{Barabsi2012} where the single knowledge items are the nodes and the links represent connections among items. A connected cluster is a corpus of knowledge. By adding knowledge three things can happen: the new knowledge item is isolated and forms an isolated cluster, it might join an existing cluster, or it may act as a connection between two clusters, fusing them together. 

Since percolation describes the patterns of linked elements under a stochastic or semi-stochastic connection mechanisms~\cite{percolationreview1,percolationreview2}, the process of filling the vector is analogous to a percolation process, and we can refer to it as the knowledge percolation problem~\cite{percolationtheory,introduction_percolation}.

The reference scenario is that of reconstructing the process that has led to the accumulation of a given corpus of knowledge, and, more important, the underlying cluster dynamics. There are many models that interpret the formation of a collaboration network by the random joining of individuals or contributions, i.e., the formation of a giant component by the establishment of random links. Our model is first of all bipartite, contributions contribute to the knowledge corpus, and the contribution overlaps gives the link among them. Moreover, we require a minimum overlap for establishing the link.

The whole corpus of knowledge can be spanned by several clusters separated by unknown elements of the corpus, or organized in a single cluster where all pieces of knowledge are connected by established relations, the process of acquiring knowledge has many similarity with the formation of a giant components in a random graph~\cite{Ding2010}.

However, in a real case, redundant links are needed for considering concepts as belonging to the same cluster or discipline. So, we assume that a new knowledge item has to have  minimum overlap with at least one of the  members already belonging to the  cluster to be inserted. 

Alternatively, this model can be seen also as a collaboration model, in which every agent knows a certain number of concepts, but is able to collaborate with others (i.e.belong to the same group), only if they have a minimum overlap (like speaking the same language and having a shared background~\cite{Fleming2007}),
evaluating therefore the possibility of agents to collaborate or to communicate with others, that could be seen as the cooperation of individuals, research groups or societies, to solve a given problem represented by the knowledge vector.

Our model can serve as an interpretation tool for examining, ex-post, how a given corpus assembled and the relative cluster dynamics. For instance, one could study how authors cluster by measuring the overlaps between citations of their papers~\cite{Newman2007}, or compare the evolution of customers of a supermarket by studying the overlap among their buying habits~\cite{Stassen1999}. Our model is however still too rough to be compared with real data.

We study how the number and size of knowledge clusters evolve when adding new items. This can be considered as a k-core growth percolation problem, although normally k-core models are studied by pruning an existing network~\cite{Kong2019}.

\begin{figure}
	\centering
	\includegraphics[width=0.99\textwidth]{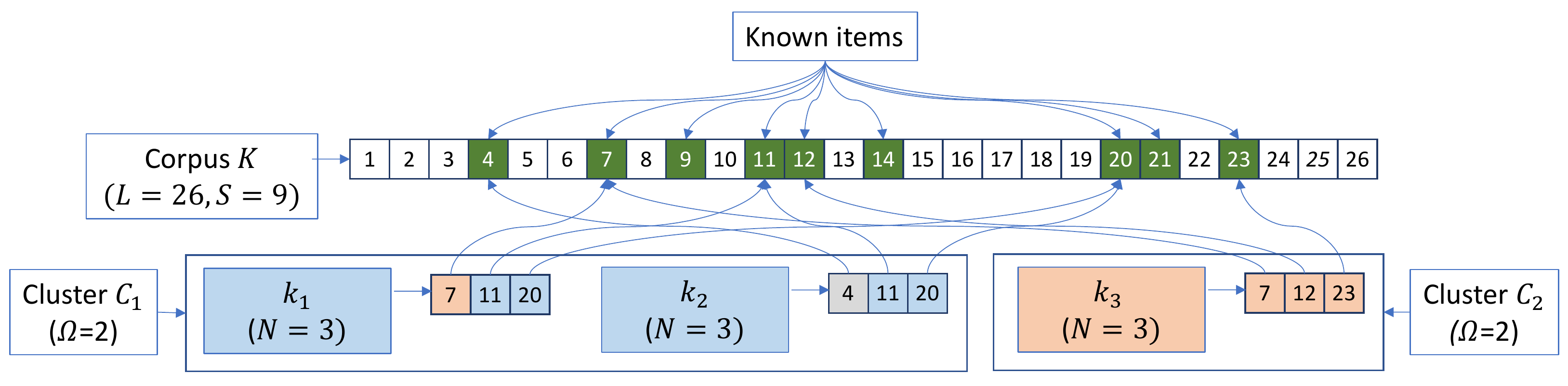}
	\caption{Schematic representation of the system in the case of a knowledge space with $L=26$ items,  $N=3$ and  $\Omega=2$ divided into two clusters $C_1=\{k_1, k_2\}$ of size $c_1=2$ and $C_2=\{k_3\}$ of size $c_2=1$. \label{fig:scheme}}
\end{figure}

\section{The model}

We  represented the corpus of knowledge $K$ as a numbered set of  $L$  items, where $K(n)=1$ if the knowledge item $n$ is present in the corpus and $K(n)=0$ if it is absent. Each contribution $k_i$ is given by a set of $N$ random items $k_i^{(n)}$, $n=1,\dots, N$ among the available ones ($k_i^{(n)}\in \{1, \dots, L\}$),. When $k_i$ is added to the corpus, we set $K(k_i^{(n)})=1$ for $n=1,\dots, N$. 

The new contribution is added to a group if it has at least an overlap of $\Omega$ to one of the elements of the group. A new contribution can also cause the fusion of two separated groups. This process is illustrated in Fig.~\ref{fig:scheme}.

Once fixed the values of $L$, $N$ and $\Omega$, the algorithm proceeds as follows:
\begin{itemize}
    \item Randomly generate a contribution $k_i$ with $N$ random items $k_i^{(n)}$ among the $L$ available ($1\le k_i^{(n)}\le  L$) without repetitions; 
    \item Add this contribution to the knowledge corpus $K(k_i^{(n)})=1$ for $n=1,\dots, N$;
    \item Check if there is any overlap with all previous contributions $k_j$  ($\forall j<i$).\\
    By denoting this overlap $\omega_{ij}=\sum_{nl} \delta_{k_i^{(n)},k_j^{(l)}}$ (where $\delta$ is the Kronecker delta), we can have three cases: 
    \begin{enumerate}
        \item $\omega_{ij}\ge\Omega$ and $k_i$  not belonging to any group:  $k_i$ is added to the cluster $C_m$ of $k_j$;
        \item $\omega_{ij}\ge\Omega$ and $k_i$ already belonging to a group: merge the cluster $C_m$  of $k_i$ and  $C_q$ of $k_j$;
        \item $\omega_{ij}<\Omega$: create a new group $C_q$ and assign $k_i$ to it.
    \end{enumerate}
\end{itemize} 

There are two different dynamics occurring in our model: how the knowledge corpus size grows, and how clusters are formed or merged together. The first one is a representation of how knowledge grows in a single individual or in a group/society, while the second one could be seen as the representation of how different branches of knowledge are intertwined~\cite{Shen2016}.

\begin{figure}
    \centering
    \includegraphics[width=0.8\textwidth]{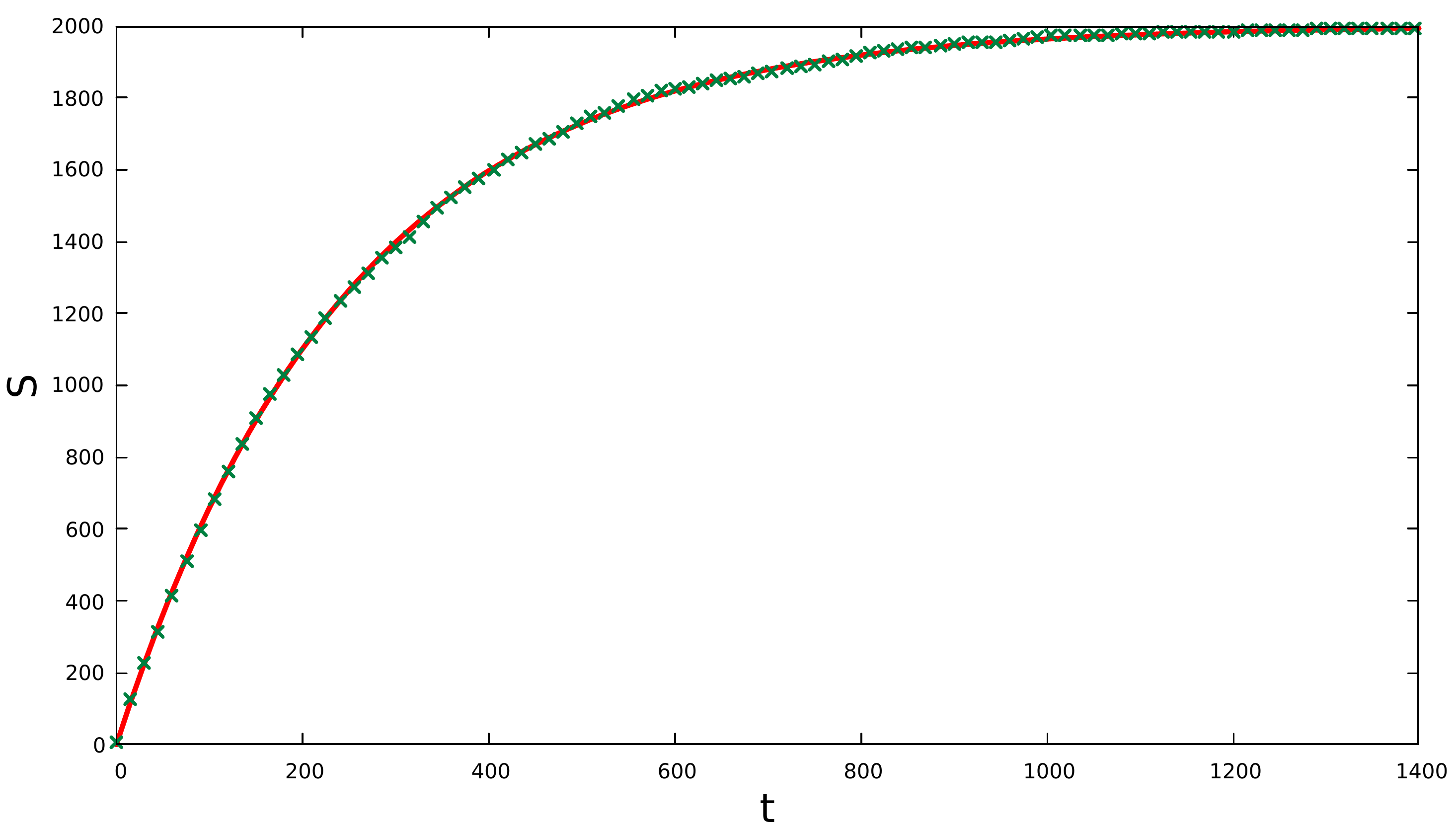}
    \caption{Knowledge size $S$ vs. time $t$ for $L=2000$, $N=8$, $\Omega=3$ (crosses) confronted with equation~\ref{eq:overlap}  (continuous line) with the same values of $N$.
        \label{fig:overlap}}
\end{figure}

\begin{figure}
	\centering
	\includegraphics[width=0.8\textwidth]{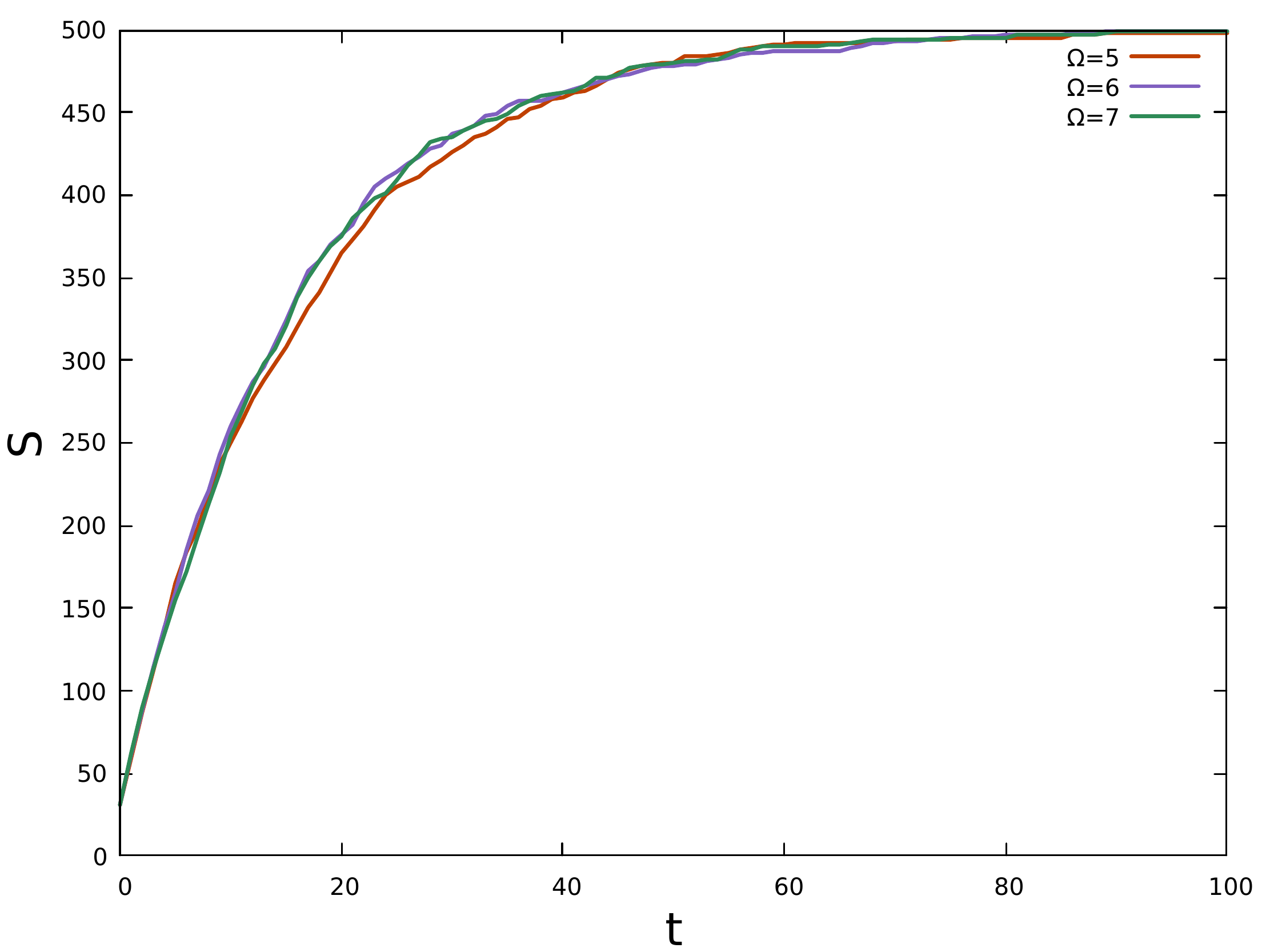}
	\caption{Knowledge $S$ vs. time $t$ with $L=500$ and $N=31$ for several values of $\Omega$. 
 \label{fig:knowledge_growth-2}}
\end{figure}

\begin{figure}
	\centering
	\includegraphics[width=0.8\textwidth]{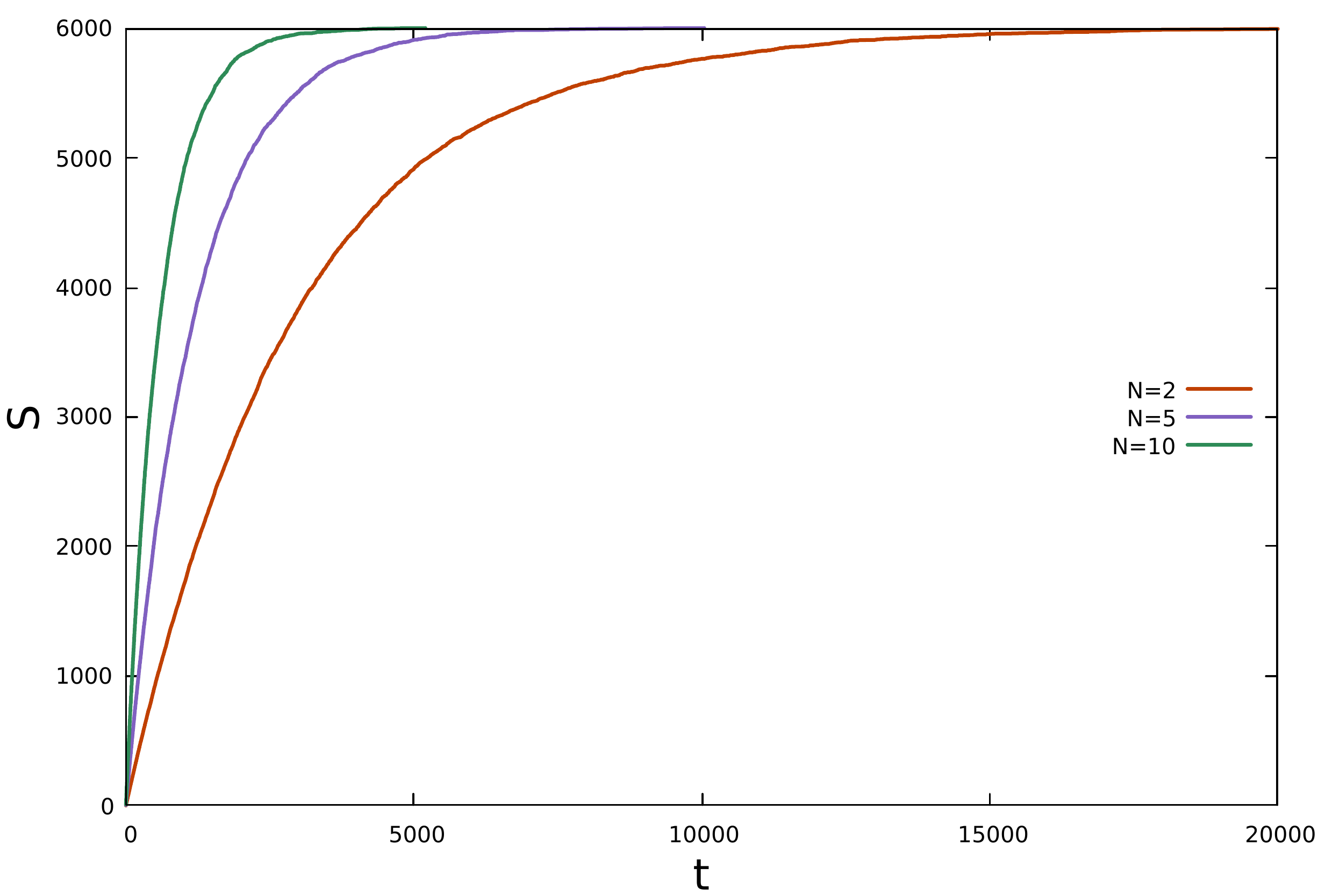}
	\caption{Knowledge $S$ vs. time $t$ with $L=6000$ and $\Omega=1$ for several values of $N$. 
 \label{fig:knowledge_growth}}
\end{figure}

\subsection{Knowledge corpus dynamics}
Let us denote the corpus size by $S=\sum_n K(n)$.
In the case  $L\gg N$, at the beginning contributions do not overlap because the probability to have overlapping items is too low. Therefore the corpus size $S$ grows linearly with time.  

The population dynamics of the corpus is an independent stochastic process, with the probability of adding an original contribution decreasing in time ($t$), while the number of those already present in the corpus ($x$) increases.

Let us examine the case with $N=1$ for simplicity.
The probability ($P(S,t)$) of having $S$ items already present at time $t$ is
\begin{equation}
    P(S,t+1)= \frac{L-(S-1)}{L} P(S-1,t) +\frac{S}{L}P(S,t).
\end{equation}
In the limit of continuous time and space, we have
\begin{equation}
\frac{\partial P}{\partial t}= - \frac{L-S}{L} \frac{\partial P}{\partial S}
\end{equation}
And using the method of characteristics~\cite{MOC} we obtain that: $P= f\left((L-S) \exp(\frac{t}{L})\right)$ and therefore the average knowledge ($\overline{S}$) grows as $\overline{S} = L (1-C\exp(-\frac{t}{L})$, 
where $C$ is an integration constant, fixed by the initial condition $\overline{S}(0)=0$, so that $C=1$, 
\begin{equation}
\overline{S} = L (1-e^{-\frac{t}{L}}) \label{average}
\end{equation}

Since the addition of knowledge is an independent process we can write Eq.~\eqref{average} with arbitrary $N$
\begin{equation} \label{eq:overlap}
\overline{S} =L(1-e^{-\frac{Nt}{L}}).
\end{equation}

Confronting Eq.~\eqref{eq:overlap} with  the results of simulations we get  an almost complete overlap as shown in Fig.~\ref{fig:overlap}, even though in simulations $\Omega > 1$. 

Indeed, it seems that the knowledge size $S$ does not depend much on the value of $\Omega$, as shown in  Fig.~\ref{fig:knowledge_growth-2}.

The knowledge size $S$ grows faster for larger values of $N$, as shown in Fig.~\ref{fig:knowledge_growth} and consistently with equation~\eqref{eq:overlap}.

\begin{figure}
    \centering
    \includegraphics[width=0.8\textwidth]{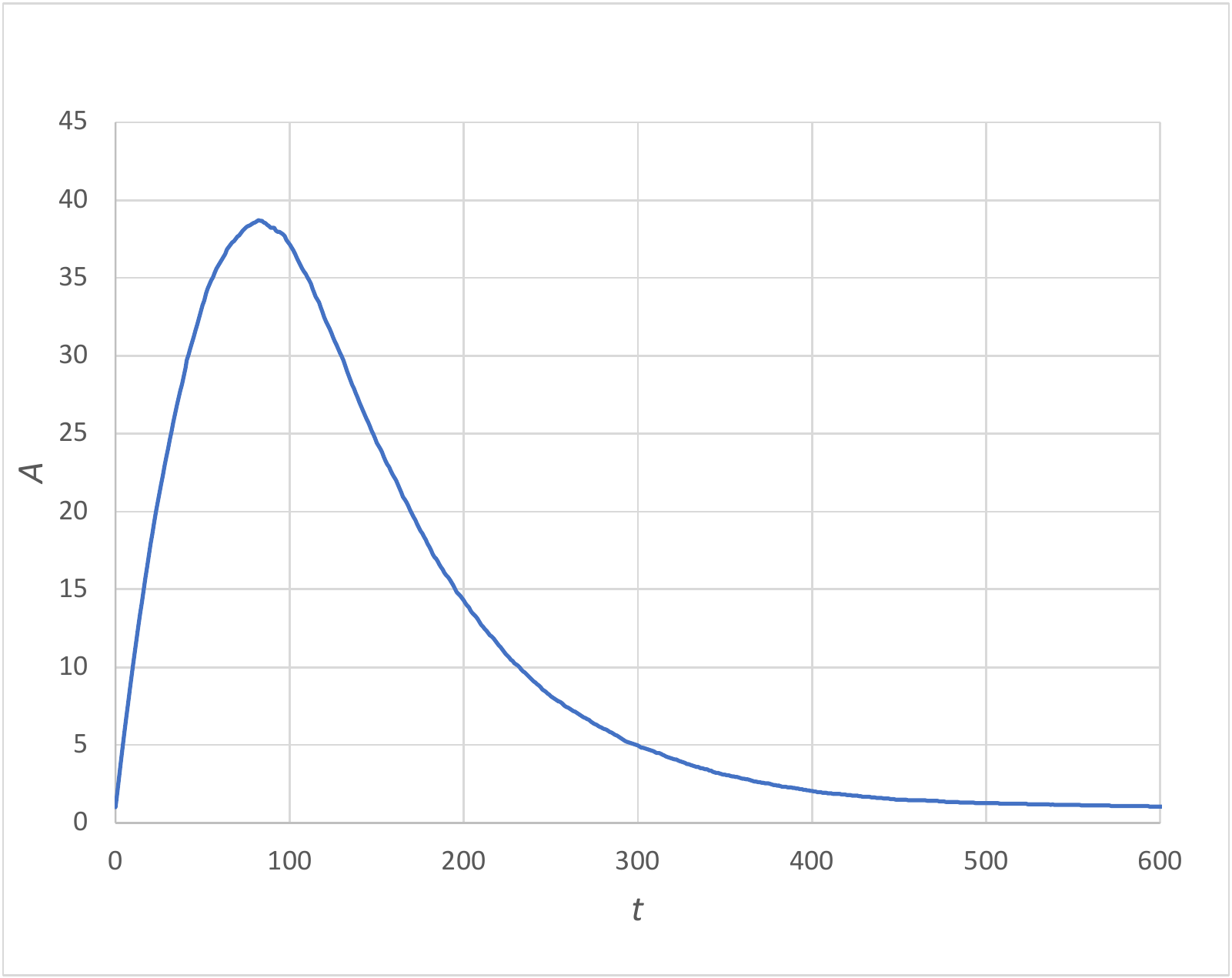}
    \caption{Number of clusters $A$ vs. time $t$ for  $L=600$, $N=2$ and  $\Omega=1$. 
        \label{fig:A}}
\end{figure}

\begin{figure}
    \centering
    \includegraphics[width=0.8\textwidth]{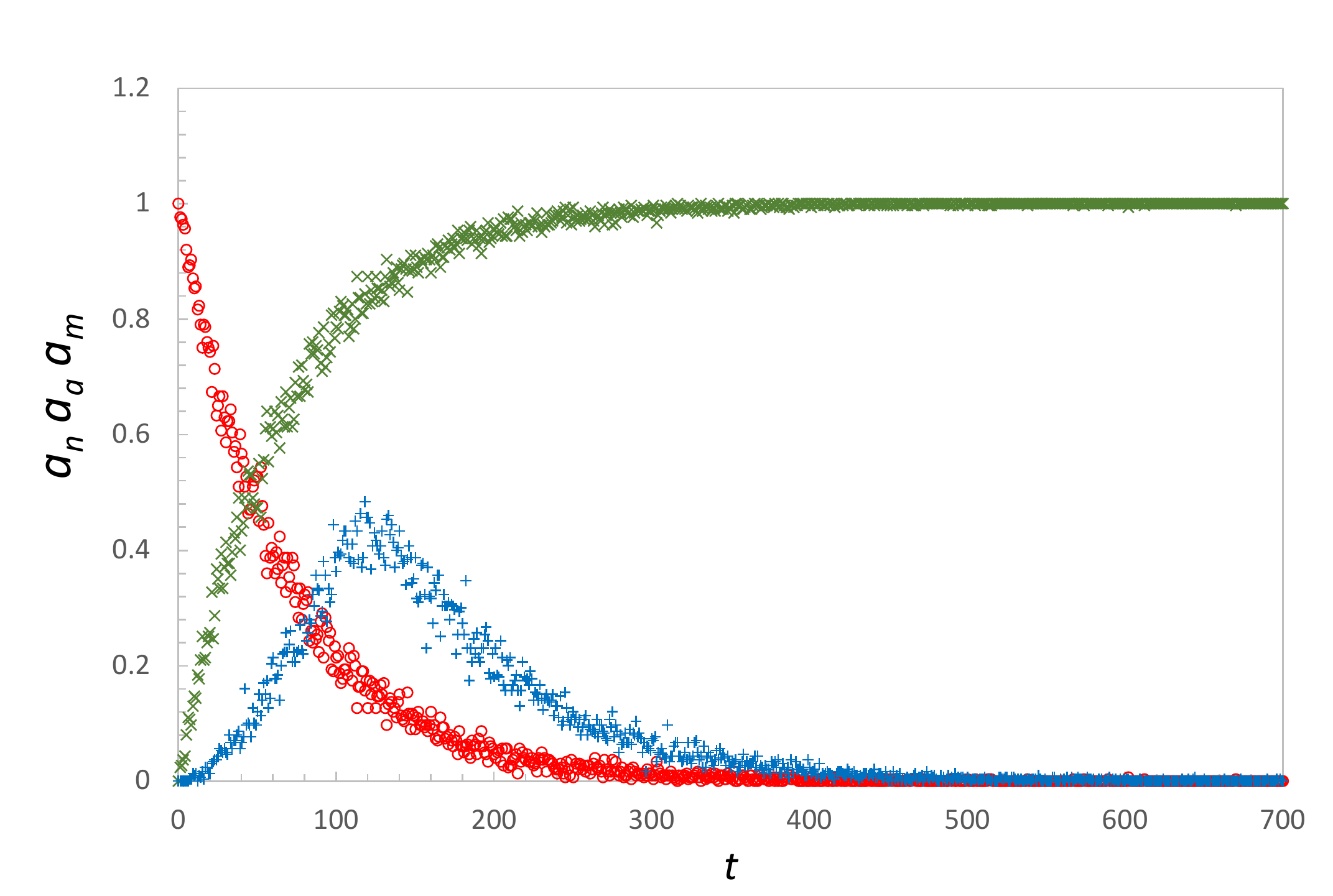}
    \caption{Number of new clusters $a_n(t)$ (red circles), number of additions $a_a(t)$ (green times signs) and number of merging $a_m(t)$ (blue pluses) vs. time for $L=600$, $N=2$ and $\Omega = 1$. 
        \label{fig:clusters}}
\end{figure}

\begin{figure}
    \centering
    \includegraphics[width=0.8\textwidth]{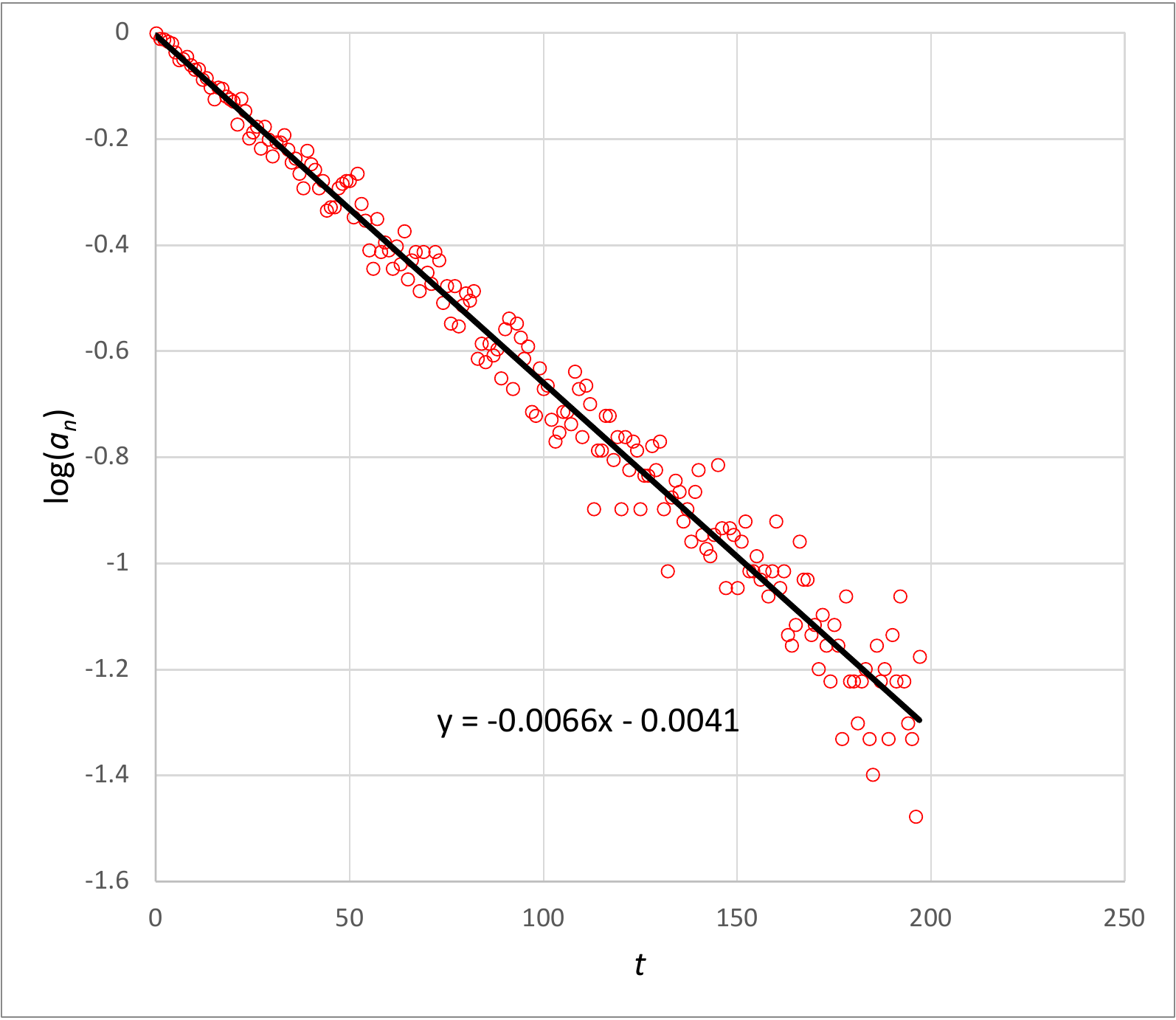}
    \caption{Number of new clusters $a_n$ vs. time $t$ in log-log scale for  $L=600$, $N=2$ and  $\Omega=1$. The exponential decreasing factor is $(3N-4)/L$.
        \label{fig:logan}}
\end{figure}

\begin{figure}
    \centering
    \includegraphics[width=0.8\textwidth]{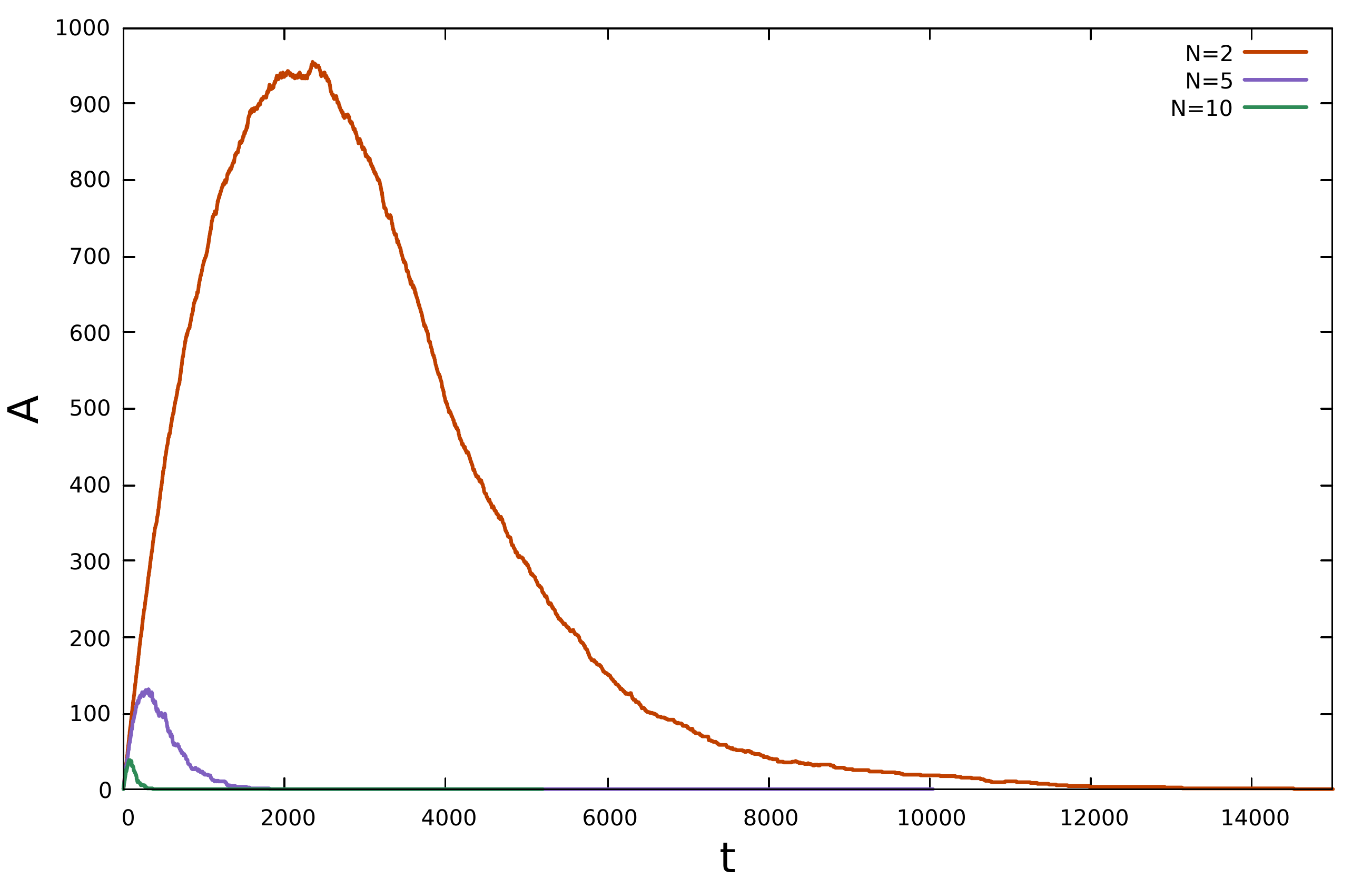}
    \caption{Number of clusters $A$ vs. time $t$ for  $L=6000$, $\Omega=1$ and varying $N$. 
        \label{fig:cluster_growth}}
\end{figure}

\begin{figure}
	\centering
	\includegraphics[width=0.8\textwidth]{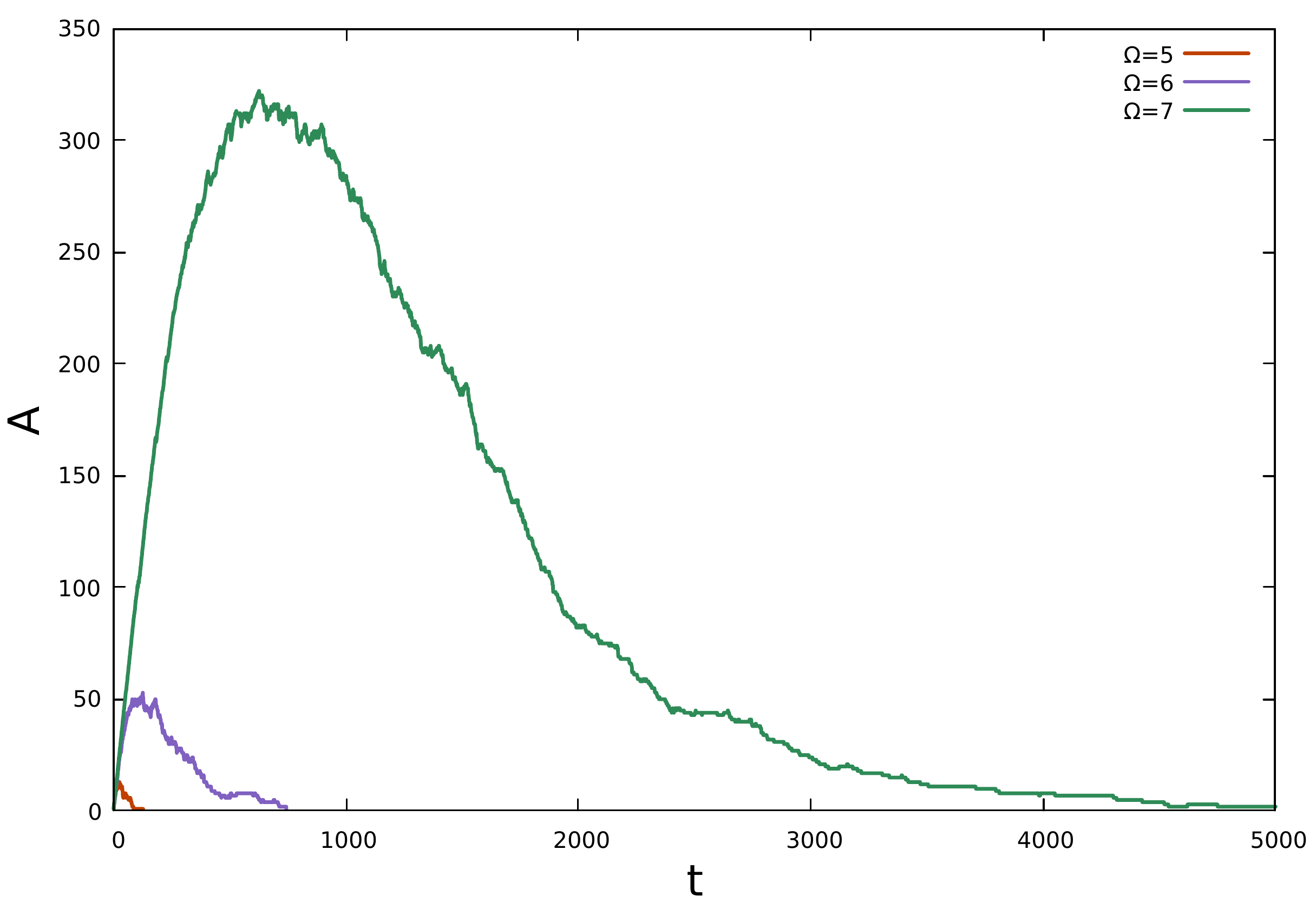}
	\caption{Number of cluster $A$ vs. time $t$ for $L=500$, $N=31$ and varying $\Omega$. 
 \label{fig:cluster_growth-2}}
\end{figure}

\subsection{Cluster dynamics}
The number of clusters $A$ at first grows with time, reaches a maximum and then decreases, ending with a single cluster, as reported in Fig.~\ref{fig:A}. 

We can distinguish three phases:
\begin{itemize}
    \item An initial almost linear growth of $A$, where   new contributions mostly form a new cluster;
    \item An intermediate phase with, in which new contributions are mostly added to an existing cluster;
    \item A decreasing behavior dominated by cluster merging.
\end{itemize}

We measured the formation of a new cluster (number of new clusters $a_n(t)$), the addition to an existing cluster (number of additions $a_a(t)$) and the merging of two clusters (number of merging $a_m(t)$), as shown in Fig.~\ref{fig:clusters}.

The actions of forming new clusters or adding it to an existing one (whether or not it causes a merging) are mutually exclusive, therefore $a_n+a_a=1$ and the total number of clusters $A(t)$ is given by
\begin{equation}
    A(t) = \sum_{\tau=1}^{t} a_n (\tau) - a_m (\tau)
\end{equation}

We can develop a simple approximation for $a_n$ in the case $N=1$, $\Omega=1$, for which we have either the formation of a new cluster (of size one) or the addition of another cluster, and no cluster merging.

By denoting with $P(A, t)$ the probability of having $A$ clusters at time $t$, we have
\[
P(A, t+1) = \frac{L-A+1}{L} P(A-1,t) +\frac{A}{L}P(A,t)
\]
which can be approximated by 
\[
\frac{\partial P}{\partial t} = \frac{A-L}{L}\frac{\partial P}{\partial A}
\]
and therefore 
\[
P=f\left((A-L)\exp\left(\frac{t}{L}\right)\right)
\]
so that, for large $A$ and small $t$, we have
\[
a_n(t) = \frac{d\overline{A}}{dt}\propto \exp\left(-\frac{t}{L}\right)
\]
which, for $N>1$ corresponds to 
\[
a_n(t) \propto \exp\left(-\frac{g(N)t}{L}\right)
\]
and numerically, as shown in Fig.~\ref{fig:logan}, we have roughly 
\[
g(N)=3N-4.
\]

\begin{figure}
    \centering
    \includegraphics[width=0.8\textwidth]{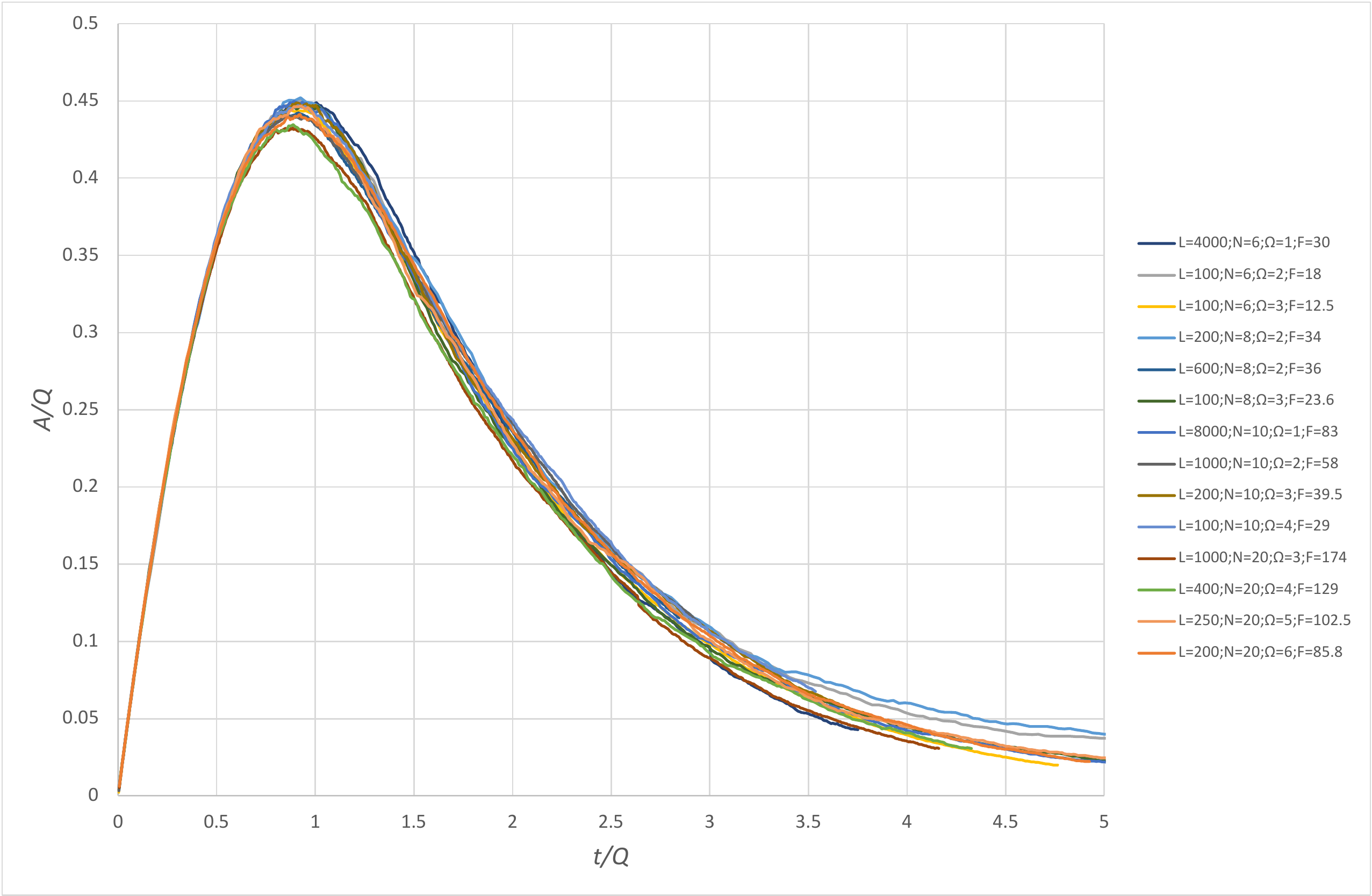}
    \caption{Scaled $A/Q$ vs $t/Q$ with $Q=(L/F)^\Omega$. 
        \label{fig:scaling}}
\end{figure}

\begin{figure}
    \centering
    \includegraphics[width=0.8\textwidth]{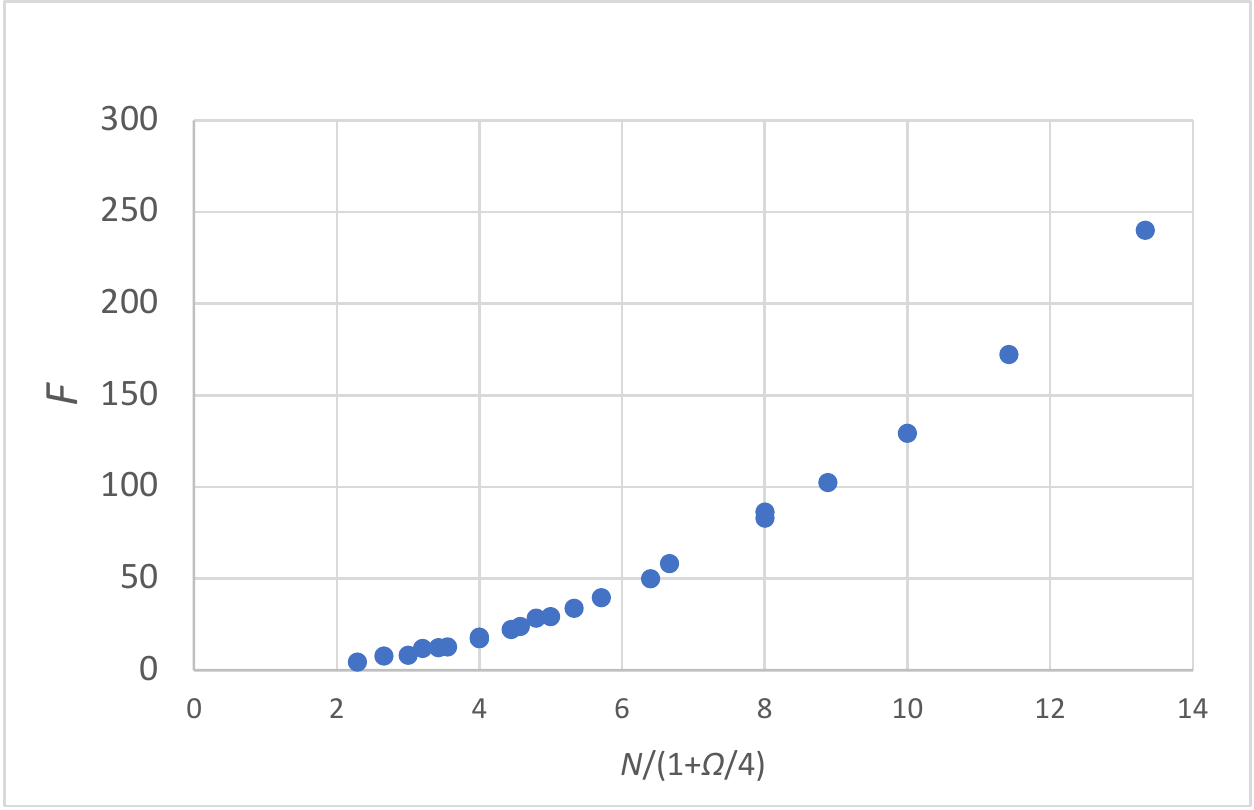}
    \caption{Scaling factor $F$ vs. $X=\frac{N}{1+\Omega/4}$. 
        \label{fig:F}}
\end{figure}

The time distribution of the numbers of clusters $A(t)$ depends on  $N$ and $\Omega$. In particular with a smaller number of items in each contribution (smaller value of $N$) is less probable to get items have an  $\Omega$  overlap and therefore a larger number of clusters $A$ will form before merging, as shown in Fig.~\ref{fig:cluster_growth}. 

On the contrary, having a smaller number of elements needed to match for merging (smaller values of $\Omega$),  means that the dynamics will reach the final state much faster, and the maximum number of clusters reached will be much smaller, as shown in Fig.~\ref{fig:cluster_growth-2}.

It is therefore expected that $A(t)$ scales with $N$ and $\Omega$. Numerically, we found that all numerical curves overlap, as shown in Fig~\ref{fig:scaling}, by rescaling $A/Q$ and $t/Q$, with 
\[
    Q= \left(\frac{L}{F}\right)^\Omega
\]
and 
\[
    F = aX^2 +bX +c
\]
with 
\[
    X=\frac{4K}{4+\Omega}
\]
as shown in Fig.~\ref{fig:F}. We have not found a valid approximation for this behavior.

\section{Conclusions}

We investigate a problem related to knowledge percolation, clustering and formation of a giant component, somewhat related to k-core percolation problems.

We studied a simple model in which each contribution is constituted by a certain number of  items, joining a cluster or even fusing two of them  when the overlap exceeds a given threshold. We showed that the growth of global knowledge and the cluster dynamics has a nontrivial time  behavior, providing some analytical approximations.

%
%

\bibliographystyle{splncs04}
\bibliography{knowledge.bib}    

\end{document}